\documentstyle[11pt,epsfig]{article}

\newlength{\dinwidth} 
\newlength{\dinmargin}
\begin{document} 
%                                                       
%======================== Report title page ===========================
%                                                                    
%            
\thispagestyle{empty}                                       
\setlength{\unitlength}{1mm}              
\begin{flushright} {\bf  Liverpool University Preprint PH 98-0901}\\
\end{flushright}                           
\begin{center}                              
                             
\vspace*{20mm}                  
{\LARGE {\bf Tagging High Energy Photons \\
 in the H1 Detector at HERA}}\\                  
\vspace{1cm}                                     

V.F.\ Andreev$^1$, A.S.\ Belousov$^1$, A.M.\ Fomenko$^1$,
L.A.\ Gorbov$^1$, T.\ Greenshaw$^2$, \\
E.I.\ Malinovski$^1$, S.J.\ Maxfield$^2$, A.V.\ Semenov$^3$, 
I.P.\ Sheviakov$^1$, P.A.\ Smirnov$^1$,\\ 
J.V.\ Soloviev$^1$, G.-G.\ Winter$^4$.

\vspace{0.5cm}

 $^1${\it Lebedev Physical Institute, Moscow, Russia.} 
\\                     
 $^2${\it Department of Physics, University of Liverpool, Liverpool, UK.}
\\ 
 $^3${\it Institute for Theoretical and Experimental 
       Physics, Moscow, Russia.}
\\
 $^4${\it DESY, Hamburg, Germany.}
%
%
%=====abstract=======        
%             
\end{center}                                                    
\vspace*{2cm}                                                          
\begin{quotation}     
\renewcommand{\baselinestretch}{1.0}\large\normalsize
\begin{center}
{\bf Abstract }
\end{center}
\vspace*{0.2cm}         
\noindent Measures taken to extend the acceptance of the H1 detector
at HERA for photoproduction events 
are described.  These will
enable the measurement of electrons 
scattered in events in the high $y$ range
$0.85 < y < 0.95$ in the 1998 and 1999 HERA run period.
The improvement is achieved 
by the installation of an electromagnetic 
calorimeter, the ET8, 
in the HERA tunnel close to the electron beam line
$8\,$m downstream of the H1 interaction point in the electron direction.
The ET8 will allow the study of tagged
$\gamma p$ interactions at centre-of-mass energies
significantly higher than those previously attainable. 
The calorimeter
design and expected performance are discussed, 
as are results obtained using a prototype placed as
close as possible to the position of the ET8
during the 1996 and 1997 HERA running. 
        
\renewcommand{\baselinestretch}{1.2}\large\normalsize                
\end{quotation}                            
%============================text==================================== 
                                                                       
\newpage                                                              
                                                                       
\section{Introduction}
The HERA accelerator at DESY brings into 
collision electrons (or positrons) with energy $27.5\,$GeV
and protons with energy $820\,$GeV ($920\,$GeV from 1998 onwards).
A large proportion of the resulting interactions can be considered
to occur between protons and essentially real
photons, that is, photons for which 
$Q^2 \sim 0$, where $Q^2 = -q^2$ and $q$ is the four-momentum 
of the photon exchanged between the electron and 
the proton. 
The energy, $E_{\gamma}$, of the photons
in these photoproduction interactions
can be related to the energies $E_e$ and $E'_e$ of the electron
in the initial and
final states through the expression
$$
y=1-\frac{E'_e}{E_e}+\frac{Q^2}{4E_e^2} 
 \simeq 1-\frac{E'_e}{E_e}
 =\frac{E_{\gamma}}{E_e}.
$$
Here $y$ is the Bjorken scaling variable defined by
$y=p \cdot q/p \cdot k$ and $p$ and $k$ are the four-vectors
of the initial proton and electron, respectively.
Measuring the energy of the outgoing electron enables the
deduction of the photon's energy; the photon is then 
said to have been ``tagged''. 
Currently, photons can be
tagged in the H1 detector over the $y$ ranges 
$0.3 < y < 0.8$ 
and  $0.08 < y < 0.15$,
as the scattered electrons then 
fall within the acceptance of the existing
electron taggers ET and ET44. These consist of
electromagnetic calorimeters located
$33.4\,$m and $44\,$m from the interaction point in the 
electron direction, the negative $z$ direction
in the HERA co-ordinate system.
 
\begin{figure}[htb]  
\centering                                                  
%\begin{picture}(285,180)
\begin{picture}(150,115)(-5,0)
   \put(120,91){\vector(0,-2){11}}  % arrow
   \put(117,92){\small\bf ET8}
   \put(100,76){\vector(0,-2){19}}  % arrow
   \put(95,77){\small\bf ET44}
\epsfig{file=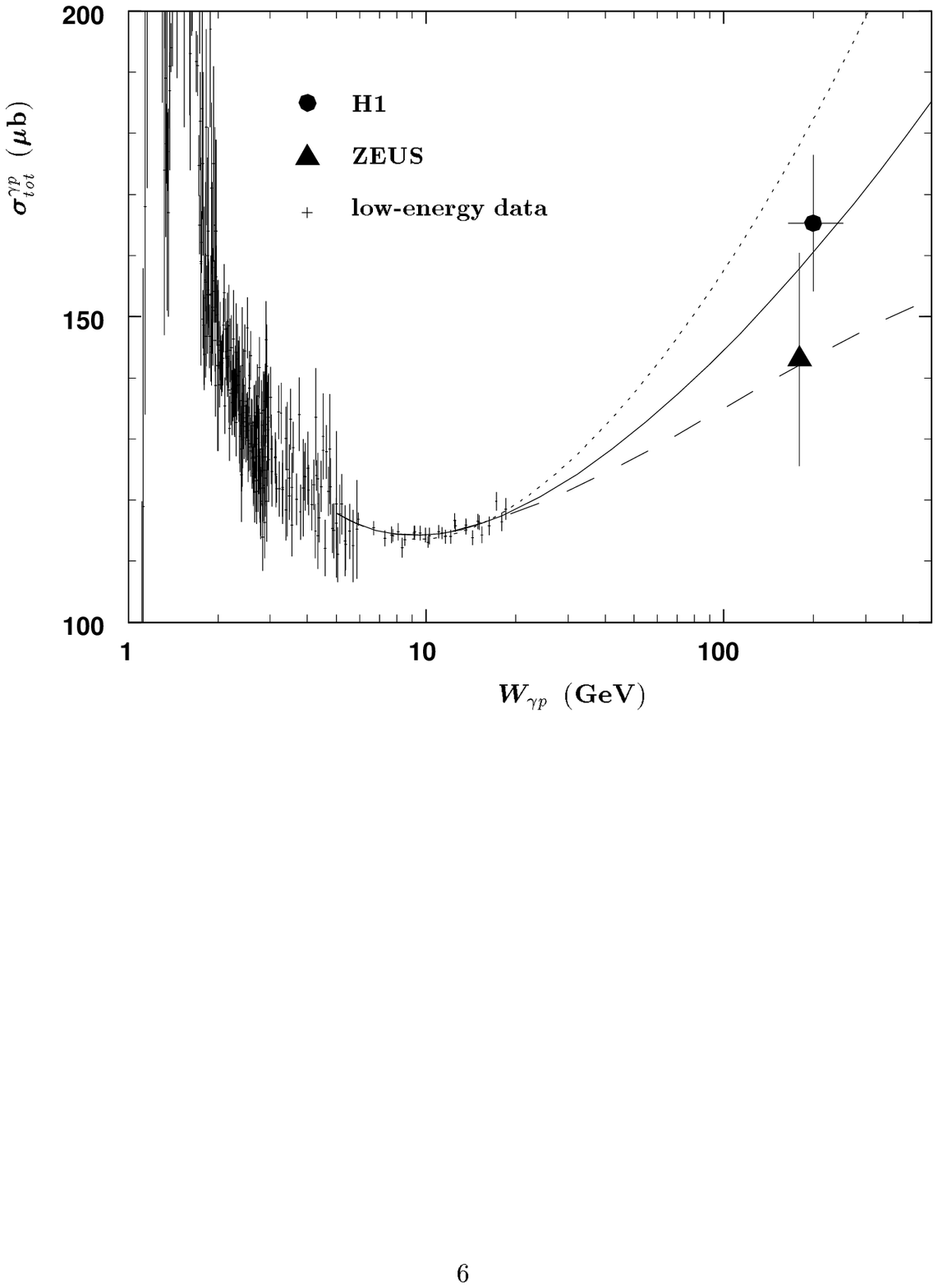,height=115mm,%
    bbllx=110pt,bblly=250pt,bburx=510pt,bbury=590pt,clip=,angle=0}
\end{picture}
\caption{\footnotesize Total photoproduction cross section as a function
         of the $\gamma p$ centre-of-mass energy W$_{\gamma p}$. 
         The arrows show the energies at which 
         measurements made using the
         ET44 and the proposed ET8 contribute ($820\,$GeV proton beam).}
\label{fig001} 
\end{figure}

This paper describes a calorimeter, the ET8, which has been
added to the electron tagger system for the 1998 and 1999 HERA running, 
and the results obtained using a prototype thereof.
The detector is located at $z \simeq -8\,$m,
enabling the detection of electrons scattered with low energy 
and hence the measurement of the scattered electron in photoproduction
interactions in which
the photon exchanged has very high energy. In addition, 
modifications to the beam-pipe have been made in this region, 
allowing the egress of the electrons through
a thin window, minimising the energy loss and scattering
they suffer before measurement.
The ET8 covers the $y$ range $0.85 < y < 0.95$. 
With the original HERA beam energies 
this extends the centre-of-mass energy
range over which photoproduction
interactions can be studied up to $W_{\gamma p} = 293\,$GeV,
the increased HERA proton beam energy 
of $920\,$GeV allows a centre-of-mass
energy of $310\,$GeV to be reached.
As is illustrated in 
figure~\ref{fig001},
%
%---------------------------- fig.1------------------- 
%---------------------------------------------------------------------- 
determination of the photon-proton cross-section
becomes possible at an energy significantly above that of the 
current highest energy results\cite{sigtot}.
In conjunction with measurements of jets in the hadronic final
state, the ET8 enables
the study of the partonic content of the 
photon in a region in which the partons carry a very small 
proportion of the photon's momentum\cite{phostruc},
corresponding to the low Bjorken-$x$ 
region which HERA has shown to be of such interest as
regards proton structure.  
The increased $y$ range over which 
photoproduction can be studied also 
allows the extraction of the energy 
dependence of various total and differential cross-sections,
such as 
$\sigma(\gamma p \rightarrow X)$, 
$\sigma(\gamma p \rightarrow J/\Psi p)$,
$\sigma(\gamma p \rightarrow J/\Psi X)$ and 
$\sigma(\gamma p \rightarrow \rho p)$.
Studies of proton structure 
will also benefit from the proposed upgrade, as one of the 
remaining uncertainties in these 
measurements\cite{H1F2,F2lowq2} 
is the contribution made 
to the proton structure data by high $y$ photoproduction 
background in which one of the hadrons is mistakenly 
identified as the scattered electron in the main H1 calorimeters.
Clearly identified photoproduction data in the
relevant region will allow this background to be quantified
more accurately than is currently possible. 

\section{Detector design}   
                                                       
\subsection{Introduction} 

In order to extend the measureable $y$
range up to $y \simeq 0.95$,
it is essential that the ET8 be able to detect
and measure accurately the energy of electrons with 
energies as low as
$1.5\,$GeV. Very little space is available
in the region in which these electrons
leave the beam pipe. The location of the 
HERA magnets dictates 
that the ET8 be placed at $z \approx -8\,$m, hence its name,
and that it be extremely compact. 
In addition, losses from the proton beam 
cause large numbers of hadronic interactions
in this region, so the ET8
must be radiation hard. 
To facilitate the separation of the signals due to electrons
from those due to hadrons, its response to both
must be well understood. 
%------------------------------ fig2.------------LUMI-system--
 \begin{figure}[htb]
 \centering
 \begin{picture}(150,70)(0,0)
  \epsfig{file=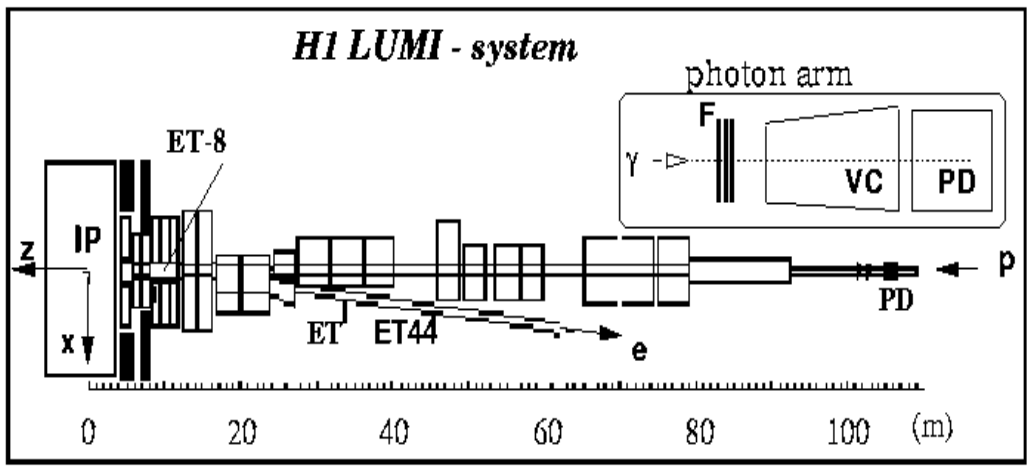,height=70mm,width=140mm,%
   bbllx=-40pt,bblly=260pt,bburx=640pt,bbury=530pt,clip=,angle=0}
\end{picture}
\caption{\footnotesize Schematic top view of the H1 luminosity system.}
\label{fig002}
\end{figure}
%----------------------------------------------------------------

The ET8 is part of the H1 luminosity system, shown in 
figure~\ref{fig002} and described briefly in the following.
A more detailed description of the H1 detector
and the luminosity system is given in reference~\cite{H1detect}.

The luminosity is determined
by measuring the rate   
of photons produced in the Bethe-Heitler process,
$e p \rightarrow e \gamma p$. Cross-checks 
are performed 
using measurements of the $e$-$\gamma$ coincidence rate,
the photons and electrons being detected
in the calorimeters of the
luminosity system. Under normal operating conditions
the axis of the photon beam coincides
with the axis of the incident electron beam and
with the $z$ axis of H1. The bremsstrahlung photons travel
inside the proton beam pipe,
leave it through an exit window at $z = -92\,$m
and hit the photon arm of the luminosity system,
the main part of which is the photon detector (PD) located
at $z =-100\,$m. A lead filter (F) and a water
\v Cerenkov veto counter (VC) are located in front of the PD. 
The filter, of $2X_{0}$
thickness, absorbs any synchrotron 
radiation and protects the photon detector 
against radiation damage. At nominal luminosity the
absorbed direct
synchrotron radiation power is about $0.4\,$kW.
The VC serves as 
an additional radiation shield of depth $1X_{0}$, 
allows rejection of events
in which a photon has converted between the exit window and
the photon detector and provides a measurement 
of some of the energy absorbed before the PD.
The scattered electrons are detected in the electron taggers
ET and ET44, located at $z=-33.4\,$m and $z=-44\,$m, respectively.
These are total absorption \v Cerenkov calorimeters 
made of KRS-15 crystals.
  
Possible choices for the ET8 
included a crystal calorimeter similar to the ET and ET44, 
a lead-scintillator sampling calorimeter
and a spaghetti type calorimeter. 
For the reasons discussed in the following
a spaghetti calorimeter is used,
as developed by H1 at DESY for the 
H1 SpaCal calorimeter~\cite{H1SpaCal}.

\subsection{The new electron tagger}

The proposed spaghetti calorimeter 
consists of BICRON BCF-12 fibres 
of $0.5\,$mm diameter
embedded in lead, with a lead to fibre ratio of $2.3:1$.
The resulting Moliere radius is $25.5\,$mm. 
The fibres emit blue light
with an emission peak near $430\,$nm.
The active volume of the calorimeter is $85\,$mm deep, 
corresponding to 10 radiation lengths. 
The calorimeter has an energy resolution that is about
$1.5$ times better than that of
the ET and ET44 crystal calorimeters.
The necessary horizontal spatial resolution is provided by a
scintillator hodoscope placed in front of the calorimeter which
allows measurement of the horizontal 
co-ordinate with an accuracy of $1\,$mm (the width of the scintillator
plates is 2 mm). The vertical co-ordinate is determined 
from the ratio
of energy deposited in the upper and lower
calorimeter modules. For the majority
of events, the electron entry point
is near the boundary between these modules, 
so the accuracy of the vertical co-ordinate determination 
is better than $1\,$mm. This spatial resolution
ensures that the energy measurement error introduced 
by corrections for leakage is less than about $5\%$.

\begin{figure}[htb]
\centering
\begin{picture}(150,200)(2,0)
\epsfig{file=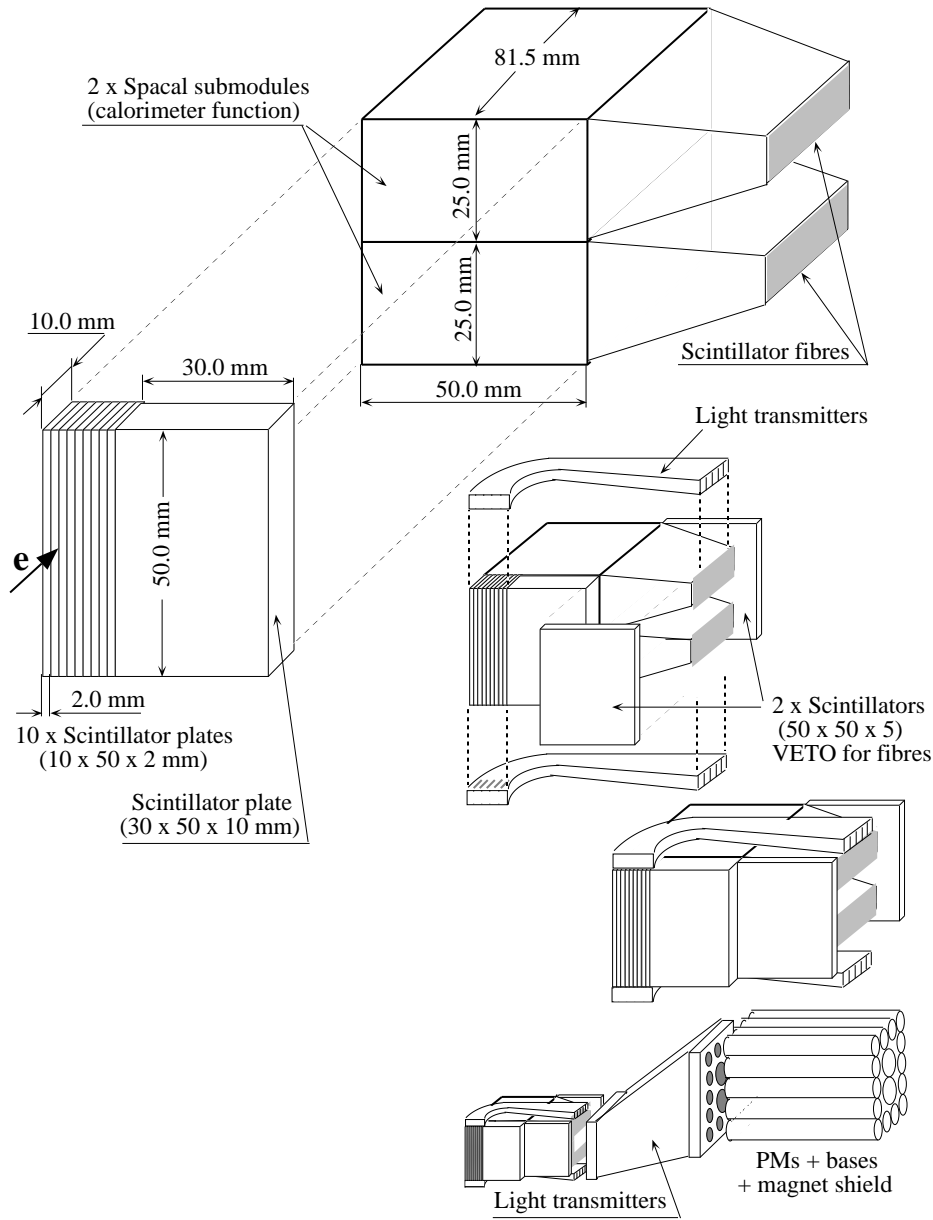,height=200mm,%
    bbllx=0pt,bblly=0pt,bburx=550pt,bbury=750pt,clip=,angle=0}
\end{picture}
\caption{\footnotesize Design of the ET8 calorimeter.}
\label{fig003}
\end{figure}

The ET8 is shown in 
figure~\ref{fig003}. A prototype has been 
built according to this design and tested as
discussed later in this paper.
Unlike the other calorimeters
in the luminosity system, the limited amount of space available
demands that the ET8
be installed on a fixed support and 
stay permanently in the HERA median plane.
The two scintillator veto plates placed before and
after the scintillator fibres allow rejection
of spurious signals due to the passage of particles
through the fibres where they exit from the lead
and are bundled to enable readout.
The intensity of \v Cerenkov light from 
particles passing through the light guides is orders of 
magnitude smaller
than the signal from electromagnetic showers in the calorimeter 
(due partly to the small lateral size of the light guide)
so no additional veto is necessary for these.
%
%---------------------------- fig.3 -------------------
%----------------------------------------------------------------

\subsection{Acceptance} 
                                                        
Acceptance studies have been performed using a fast
simulation package, H1LUMI ~\cite{LEVO}.
This models the current HERA electron ring optics
as described, for example, in~\cite{BRINK}.
Several possible positions for the new tagger were considered. 
%
%---------------------------- fig.4 -------------------
\begin{figure}[htb]
\centering
\begin{picture}(120,100)(-5,-7)
\epsfig{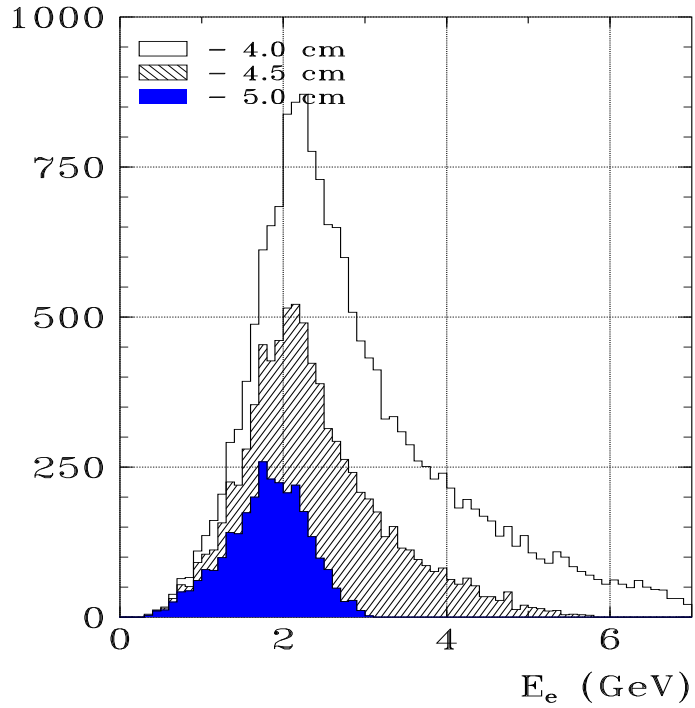}
\end{picture}
\caption{\footnotesize Energy distribution of the scattered 
         electrons in ET8
         as a function of distance from the $e$-beam axis. } 
\label{fig004}
\end{figure}
%----------------------------------------------------------------------
%
After taking into account the
apparatus currently present in the HERA tunnel,
the space needed for
the calorimeter itself and the access requirements for
installation and surveying,
the position at $z \simeq -8\,$m was chosen. Small possible variations
of the beam parameters at the IP (horizontal and vertical tilt
of $\pm 0.15$ mrad and offset of $\pm 1\,$mm) were taken into
account in the simulation. They
typically lead to $15$ to $25\%$
changes in the electron tagger acceptance.
%                              
%---------------------------- fig.5-------------------
\begin{figure}[htb]
\vspace{0.5cm}
\centering
\begin{picture}(150,90)(-5,-7)
\epsfig{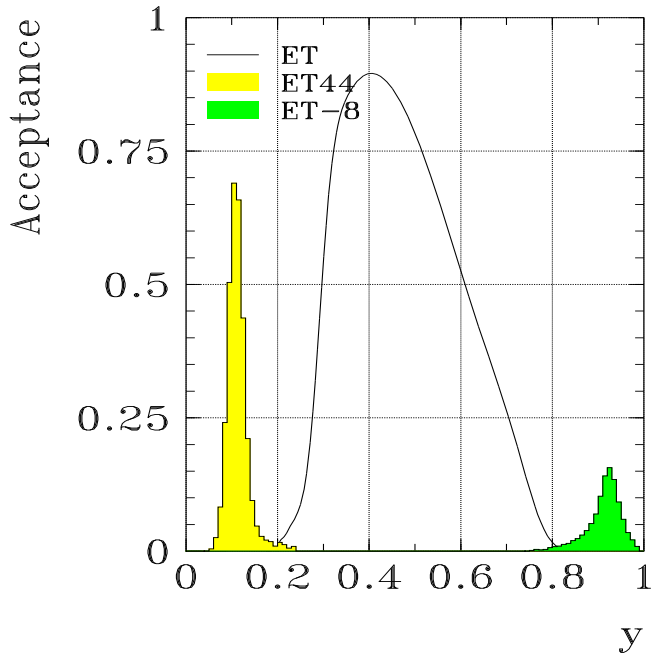}
\end{picture}
\caption{\footnotesize The acceptance for photoproduction events
         of the three electron taggers 
         with nominal beam conditions.}
\label{fig005}
\end{figure}
%----------------------------------------------------------
%
Variations of $\pm 1\,$mm in the detector
position give an uncertainty of
about $15\%$ in the ET8 acceptance.
Hence, precise knowledge of the both the 
tagger position with respect to
the electron beam and of the beam parameters
is very important.
The acceptance can
be defined with high precision from the data using 
$ep \rightarrow e \gamma p $ events, as is done for
the present ET and ET44 calorimeters. This reduces
uncertainties related to the precision with which real beam
conditions can be simulated.
%
%---------------------------- fig.6-------------------
\begin{figure}[htb]
\centering
\begin{picture}(150,98)(0,-7)
\epsfig{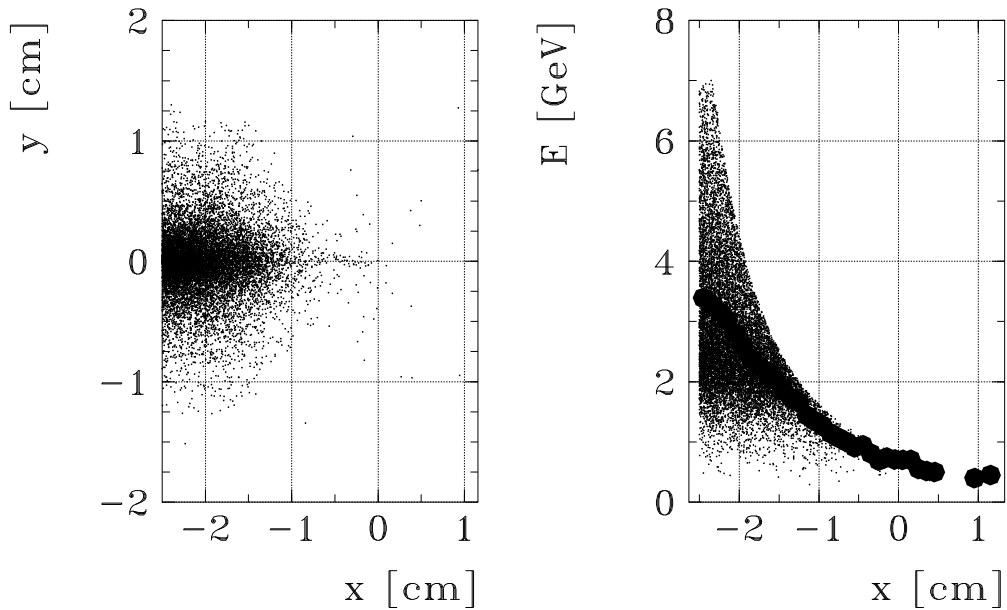}
\end{picture}
%    bbllx=0pt,bblly=0pt,bburx=470pt,bbury=800pt,clip=,angle=90}
\caption{\footnotesize Distribution of co-ordinates of 
         point at which electron enters ET8 (left) and correlation 
         between horizontal co-ordinate of entry point
         and energy (right) in $\gamma p$ events. 
         Filled circles represent average
         values with one sigma error bars. The $x$ co-ordinate 
         is measured from the centre of the ET8.}
\label{fig006}
\end{figure}
%----------------------------------------------------------------------
%
The energy distribution of the electrons accepted by the ET8
as a function of distance from the electron beam axis
is shown in figure~\ref{fig004}.
Figure~\ref{fig005} shows the acceptances of all the 
luminosity system electron detectors
(ET, ET44 and ET8) for
photoproduction events with nominal beam conditions (zero tilt and 
offset). The ET8 is seen to provide
the desired extension of the  
measureable $y$ range to the high $y$ interval
$0.85 < y < 0.95$.

The distribution of the
horizontal co-ordinate of the point at which the electron 
enters the ET8 is shown in figure~\ref{fig006}. 
The lateral size of the calorimeter adequately
covers this range.
In addition, figure~\ref{fig006} shows the 
correlation between the measured electron energy
and the point at 
which it enters the detector.
This illustrates how precise co-ordinate measurement
can aid calibration of the ET8 and 
reconstruction of the electron energy, 
in addition to allowing accurate corrections for
shower leakage.

\subsection{Radiation hardness of materials}      

The H1 luminosity system, which is close to the 
HERA beam lines, suffers
a considerable radiation load. For example, 
the PD typically receives a 
dose of $\approx 10\,$Mrad during a year's HERA running.
Therefore the study of the radiation hardness of 
the materials
(veto scintillators, fibres, plexiglass) used in
the ET8 calorimeter is an important task. The influence of
radiation on the materials used was studied using
the bremsstrahlung photon beam at the
$1\,$GeV electron synchrotron of the Lebedev Physical 
Institute in Moscow.
%---------------------------Fig.7-----------
\begin{figure}[htb]
\centering
\begin{picture}(175,100)(0,0)
\put(155,85){\Large \bf a}
\put(155,35){\Large \bf b}
\epsfig{file=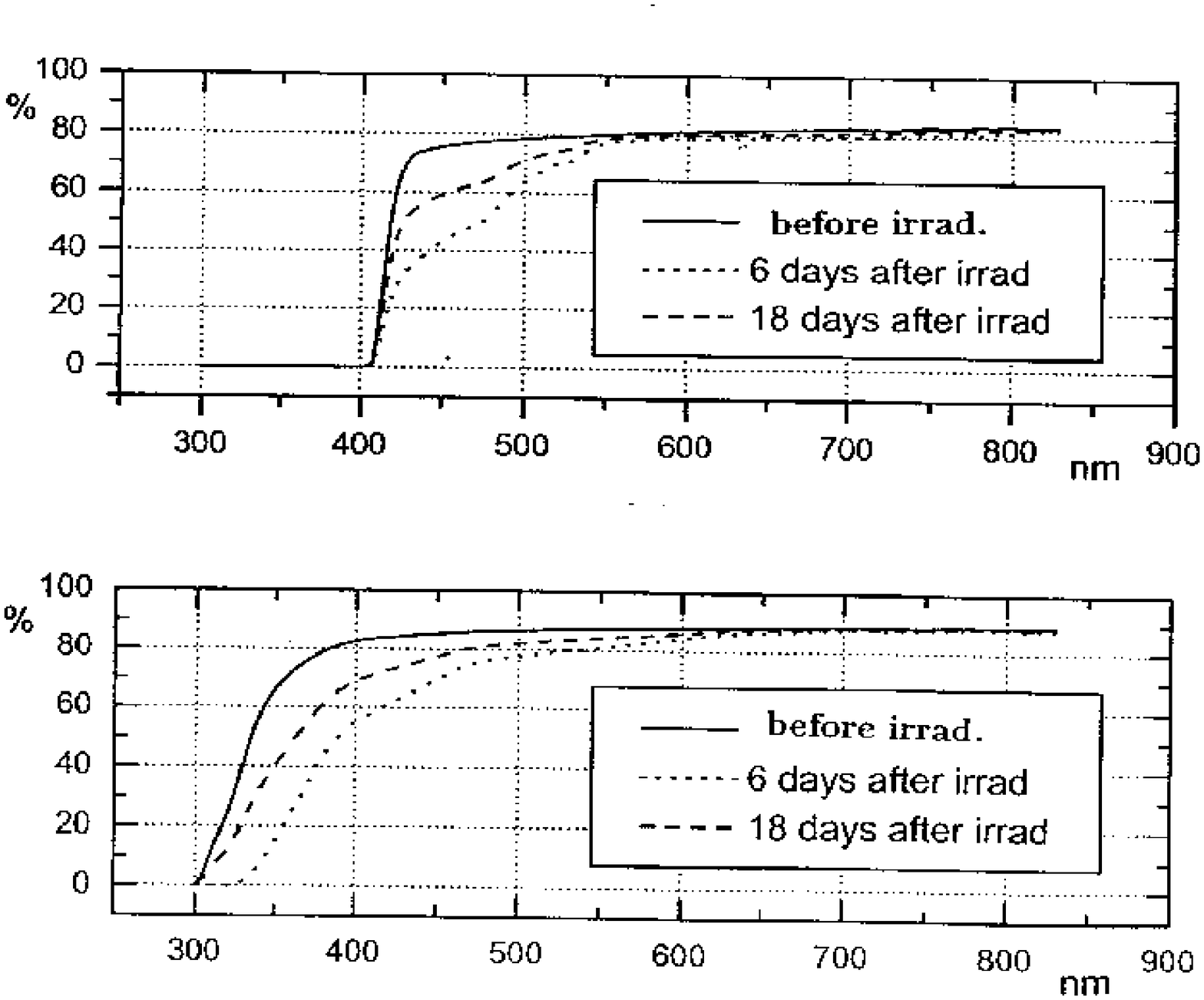,height=100mm,width=145mm,%
    bbllx=0pt,bblly=130pt,bburx=620pt,bbury=670pt,clip=,angle=0}
\end{picture} 
\caption{\footnotesize Scintillator (a) and plexiglass 
         (b) optical transmission
         as a function of wave length $\lambda$.}
\label{fig007}
\end{figure}
%--------------------------------------------------- 
Samples of the materials to be studied were placed at the 
beam collimator of the bremsstrahlung
beam which is derived from the collision of 
the accelerated electrons with an internal
target. Dose determinations were made using 
dosimeters adjacent to the samples. 
Measurements of
the change of the attenuation length of the BICRON fibres
were made using the procedure described in~\cite{BARAN}. 
The total dose administered to the fibres
was $1\,$Mrad.
The measurements show that the photon mean free path 
decreased from about $130\,$cm to about $60\,$cm after
irradiation. As the expected yearly 
dose at the position of the ET8
due to nominal luminosity HERA running is 
about $0.1\,$Mrad, the functioning of the calorimeter will
not be seriously impaired by the radiation absorbed
by the fibres.  

The veto scintillator and the plexiglass used as
light guides in the ET8 received a total
dose of about $2\,$Mrad. This caused visible yellowing
of the scintillator, whereas the
plexiglas became brown. 
The transparency of the
samples before and after irradiation 
in the wavelength range $300$ to $800\,$nm was
determined using a spectrophotometer. The results are shown
in figure~\ref{fig007}. The changes due to the 
irradiation are seen to be essentially in the
short wavelength region. Several days after irradiation the optical
transparency was observed to be partially restored. 
  
In conclusion, the tests show that the decrease of the light yield
due to the radiation damage suffered by the components of the ET8
will not influence the detector response
significantly. The ET8 will be continuously calibrated,
as are all the luminosity system detectors. This procedure
will correct for the small losses expected under 
normal HERA operating conditions.

\section{Prototype test results}

\subsection{Design of prototype} 
 
A prototype ET8 calorimeter was built according to the  
design shown in figure~\ref{fig003}.
In order to test the prototype ``in situ'', it
was installed in the tunnel as close to the proposed position 
of the ET8 as was possible.
The beam line configuration in 1996 and 1997
required that the prototype ET8 be placed at $z=-7\,$m, hence it 
was dubbed the ET7. Note also that the 1996 and 1997 running 
was done with positron rather than electron beams.

The level 1 trigger element for the ET7
was the simple condition $E_{tot} > E_{thr}$, where
$E_{tot}$ is the sum of the energy deposited in all the
calorimeter modules
and $E_{thr}$ the threshold energy, which must 
have a value as low as $E_{thr} \simeq 0.5\dots1.0\,$GeV.
Note that a minimum ionising particle loses about $50\,$MeV of
energy in traversing the ET7. 
An additional $\gamma$-veto condition from the photon 
arm of the luminosity system
suppressed the high rate
background from the Bethe-Heitler process 
with an efficiency of about $98\%$.

\subsection{Event selection and rate} 

%-------------------------- fig.8----------- 
\begin{figure}[htb]
\centering
\begin{picture}(170,90)(5,0)
\epsfig{file=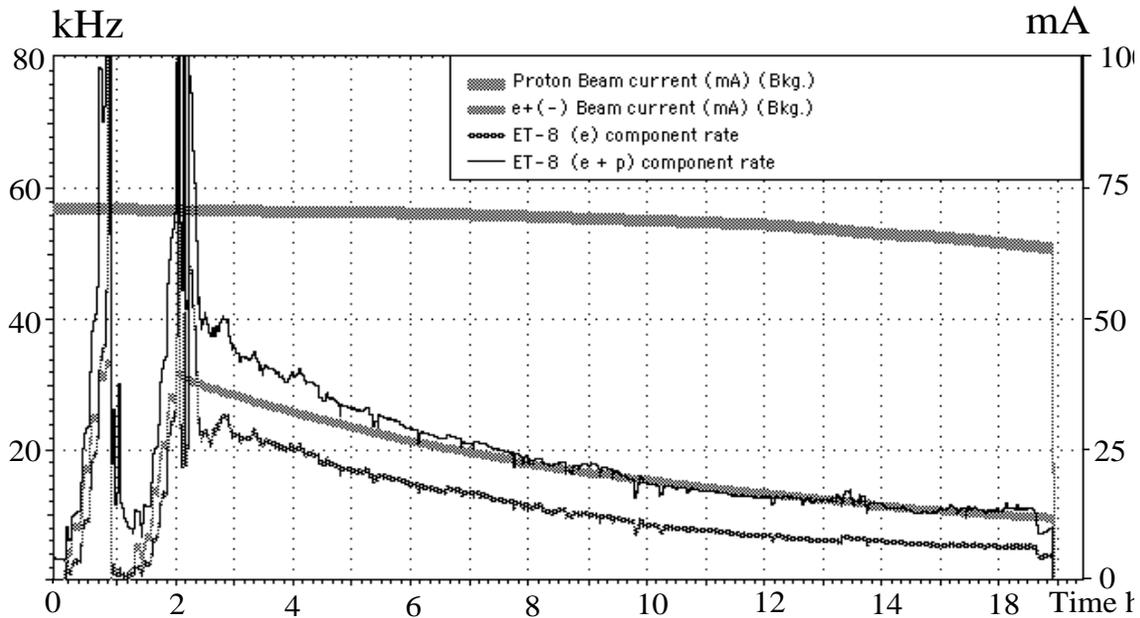,height=90mm,width=160mm,%
    bbllx=0pt,bblly=0pt,bburx=670pt,bbury=400pt,clip=,angle=0}
\end{picture}
\caption{\footnotesize Positron and proton beam 
         currents as a function of time
         during a luminosity run with the measured 
         rates of ET7 calorimeter triggers with ($e$) and 
         without ($e+p$) the requirement that 
         the signal be temporally
         associated with the electron signal.
         (Note, ET7 labelled ET-8 in figure)}
\label{fig008}
\end{figure} 
%---------------------------------------------------
The rate of events observed by 
the ET7 during one luminosity
run is presented in figure~\ref{fig008}. 
The rate spectrum shows spikes caused by particle losses 
during the injection of the 
proton and positron beams.
Apart from these, the rates are seen to 
be roughly proportional to the 
product of the positron and proton beam currents,
suggesting that the observed events are 
predominantly due to positron-proton collisions.  

\begin{figure}[htb]
\centering
\begin{picture}(160,80)(-5,-90)
\epsfig{file=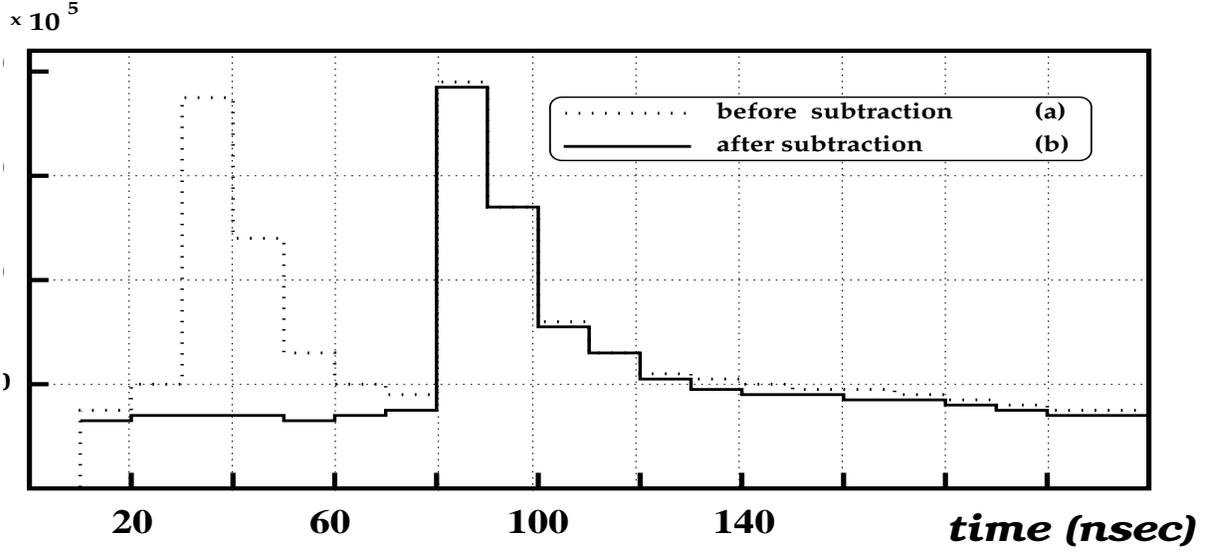,height=160mm,width=110mm,%
    bbllx=0pt,bblly=20pt,bburx=520pt,bbury=640pt,clip=,angle=-90}
\end{picture}
\caption{\footnotesize The averaged FADC pulse 
         shape for the events triggered by
         the prototype ET7 subtrigger (a) with proton halo and
         (b) after subtraction of proton halo.}
\label{fig009}
\end{figure}

The time structure of the ET7 signals was 
measured using flash analogue to digital converters.
The averaged FADC pulse shape of the calorimeter's
response for the
events triggered by the 
ET7 subtrigger is shown in
figure~\ref{fig009}. Two peaks are seen at different times.
The earlier of these 
is due to the collisons of particles from the proton 
beam halo with the ET7 and the later to 
scattered positrons, the time between the two being
$54\,$nsecs as expected given the distance between the
interaction point and the ET7.
This time difference can be used to 
isolate the positron signal by defining a ``positron 
time window'' about the expected positron time of arrival. 
%-------------------------- Fig.9 ----------- 
%---------------------------------------------------
This was used to refine the 
ET7 level one trigger. 
For the studies described in the following it 
was required that $E_{thr} = 0.3\,$GeV
and that the 
signal from the ET7 be in the positron time window.

\subsection{Detector calibration}

In common with the other detectors in the luminosity 
system, $ep \rightarrow e\gamma p$ events are used
for energy calibration purposes. Since the energy 
transfer to the proton is negligible, the relation  
$$
  E'_e + E_{\gamma} = E_e
$$
must hold. The absolute energy scale of the PD 
can be determined using the high energy edge of the 
bremsstrahlung spectrum after which the
energy detected in the PD, $E_{PD}$, provides an 
estimate of $E_{\gamma}$.
Studies of the
summed energy in the photon arm of the luminosity
system and the relevant electron
tagger then allow calibration of the latter~\cite{H1ISR}.
  
The calibration of the ET7 proceeds by selecting
events which have a positron impact
point on the calorimeter's front plane  
well removed from its edges. 
The calibration constants, $C(n)$, of the $N$ ET7 channels 
are then determined by minimising the expression
$$
\sum_{i} \Big(\sum_{n=1}^{N} C(n) A_{i}(n) + E_{PD} - E_{e}\Big)^{2},
$$
where 
$A_{i}(n)$ is the amplitude of the signal from the
$n^{{\rm th}}$ channel in the
$i^{{\rm th}}$ event.
 %-------------------------- fig.10----------- 
\begin{figure}[htb]
\centering
\begin{picture}(170,90)(0,-80)
\epsfig{file=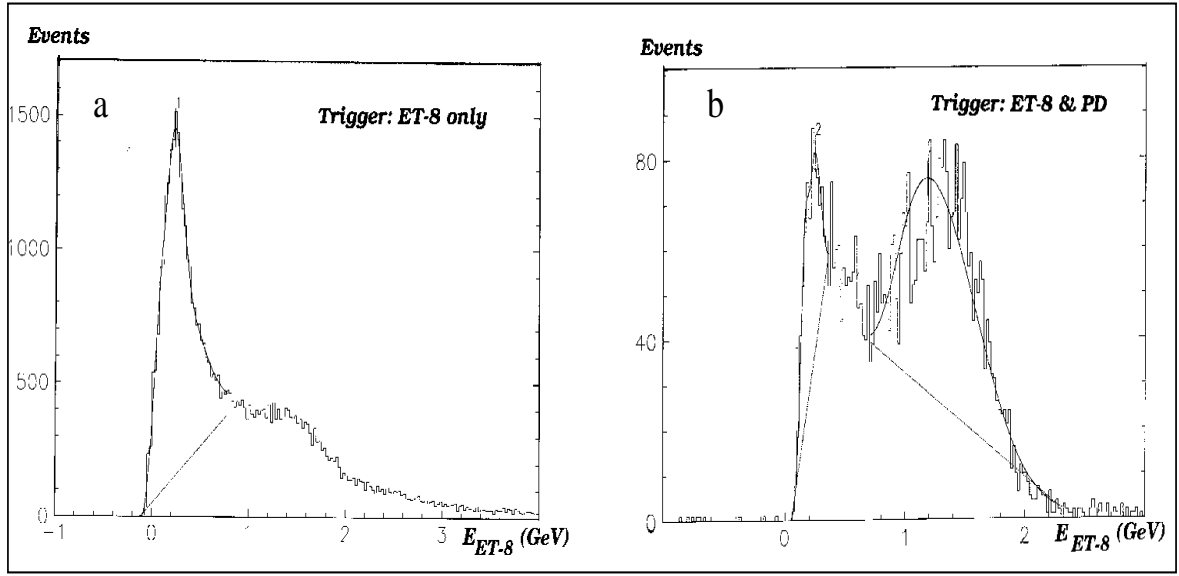,height=160mm,width=80mm,%
    bbllx=200pt,bblly=140pt,bburx=410pt,bbury=650pt,clip=,angle=-90}
\end{picture}
\caption{\footnotesize Absolute energy distribution 
     from the prototype ET7 
     after (a) level one trigger, (b) demanding coincidence
     with a signal from the photon detector. (Note, ET7 
     labelled ET-8 in figure.)}
\label{fig010}
\end{figure} 
%---------------------------------------------------
Using this calibration gives
the absolute energy spectrum, measured in the positron
time window, shown in
figure~\ref{fig010}a. The distribution
displays the expected 
peak between $1$ and $2\,$GeV, corresponding to
events with $E_{\gamma} \approx 26\,$GeV, that is, $y \approx 0.9$. 
There is another large 
contribution to the distribution 
which is strongly peaked at smaller energies.
This is mainly due to 
background arising from 
off-momentum positrons accompanying the beam. 
This can be suppressed by 
demanding the coincidence of the ET7 signal 
with a signal from the PD as this
selects events from the $ep \rightarrow ep\gamma$
process. 
The energy
distribution of positrons 
following this selection is 
presented in figure~\ref{fig010}b. 
There still remains a significant tail to low energy.
This is due to the small angle at which the 
detected positrons traversed the beam-pipe.
They travelled through a significant amount of material,
often initiating electromagnetic showers and hence depositing
only a fraction of their energy in the ET7. 
This has now been remedied by the installation 
of a modified beam-pipe with the necessary thin exit window.
 %-------------------------- fig.11----------- 
\begin{figure}[htb]
\centering
\begin{picture}(170,90)(0,0)
\epsfig{file=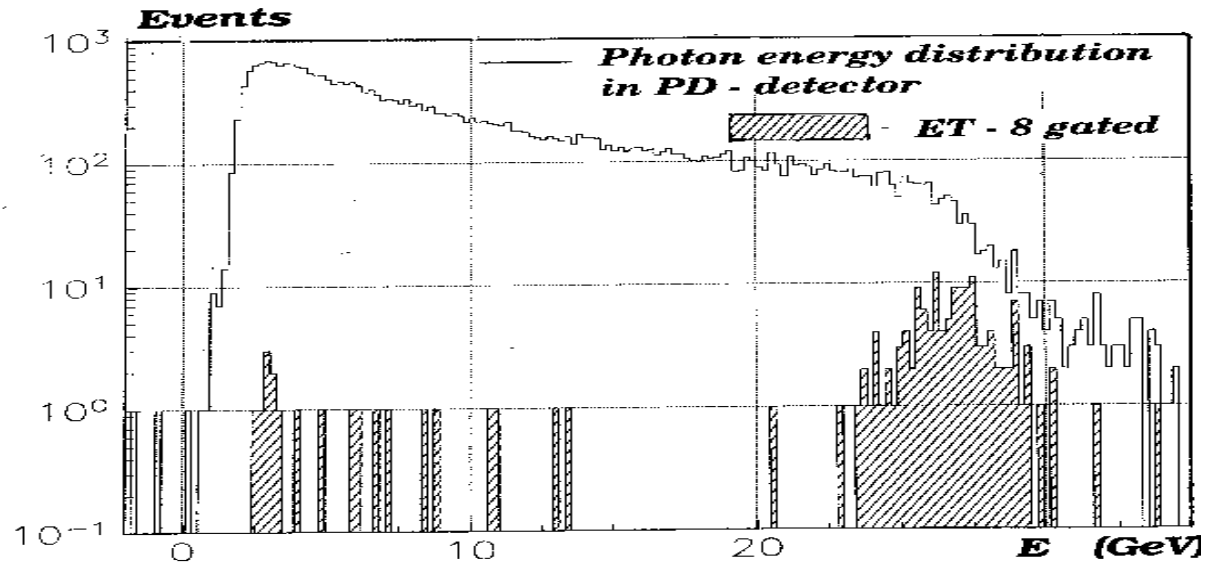,height=80mm,width=160mm,%
    bbllx=20pt,bblly=200pt,bburx=585pt,bbury=590pt,clip=,angle=0}
\end{picture}
\caption{\footnotesize Photon energy spectra measured by the PD:
      solid line, all measured events; shaded area, 
      events triggered by ET7. (Note, ET7 
      labelled ET-8 in figure.)}      
\label{fig011}
\end{figure} 
%---------------------------------------------------

Figure~\ref{fig011} shows the measured energy spectrum of all 
Bethe-Heitler photons seen in the 
PD. The shaded
area represents events in which the positron was detected in the 
ET7.
%-------------------------- fig.12----------- 
\begin{figure}[htb]
\centering
\begin{picture}(120,80)(-15,0)
\epsfig{file=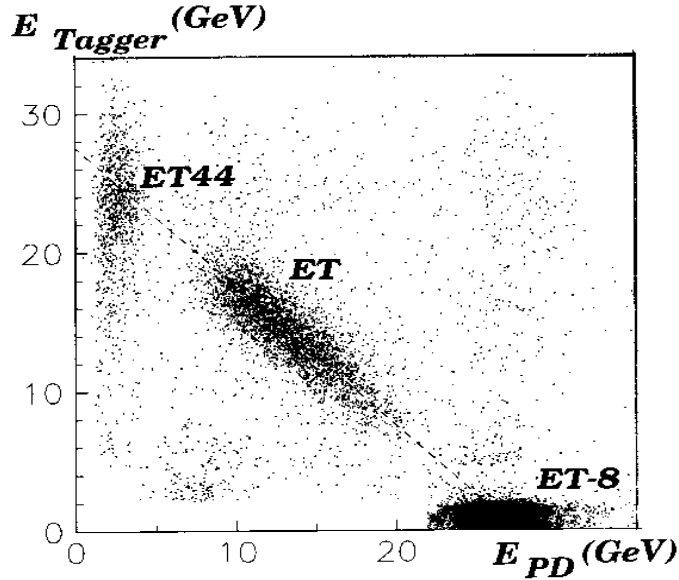,height=80mm,width=90mm,%
    bbllx=65pt,bblly=150pt,bburx=550pt,bbury=650pt,clip=,angle=0}
\end{picture}
\caption{\footnotesize $E_{\gamma}$ against $E_{e}$ for bremsstrahlung
     events detected by the H1 luminosity system including 
     electron taggers ET, ET44 and ET7. (Note, ET7 
     labelled ET-8 in figure.)}
\label{fig012}
\end{figure} 
%---------------------------------------------------
%
Note that the 
showering in the beam-pipe which occurs in front of the ET7
does not affect this spectrum, and the signal is indeed
observed to be very clean.
The mean value of this distribution is about $26.2\,$GeV,
which compares well with the expected average energy 
of bremsstrahlung photons obtained
from MC simulations including the effects of the 
acceptance of the ET7. 

Figure~\ref{fig012} illustrates the energy correlation between 
$E_{\gamma}$ and $E_{tag}$ for Bethe-Heitler events
selected by requiring the coincidence of
signals from any one of the ET, the ET44 or the ET7 
and the PD. 
Due to the large differences in the trigger 
rates of the different taggers,
(ET44 $\simeq 280\,$kHz, ET $\simeq 170\,$kHz and
ET7 $\simeq 15\,$kHz at a luminosity ${\cal L} = 5 \times 
10^{30}\,$cm$^{-2}$s$^{-1}$) these data were taken
with differing pre-scale conditions for the trigger elements
involving the different electron
taggers. The anti-correlation 
arising from the requirement that $E_{\gamma} + E'_e = E_e$ can
clearly be seen. The above demonstrate that the 
ET7 was indeed detecting and measuring the energy of positrons
scattered in $ep \rightarrow ep\gamma$ interactions.

\begin{figure}[htb]
\centering
\begin{picture}(160,110)(0,-90)
\epsfig{file=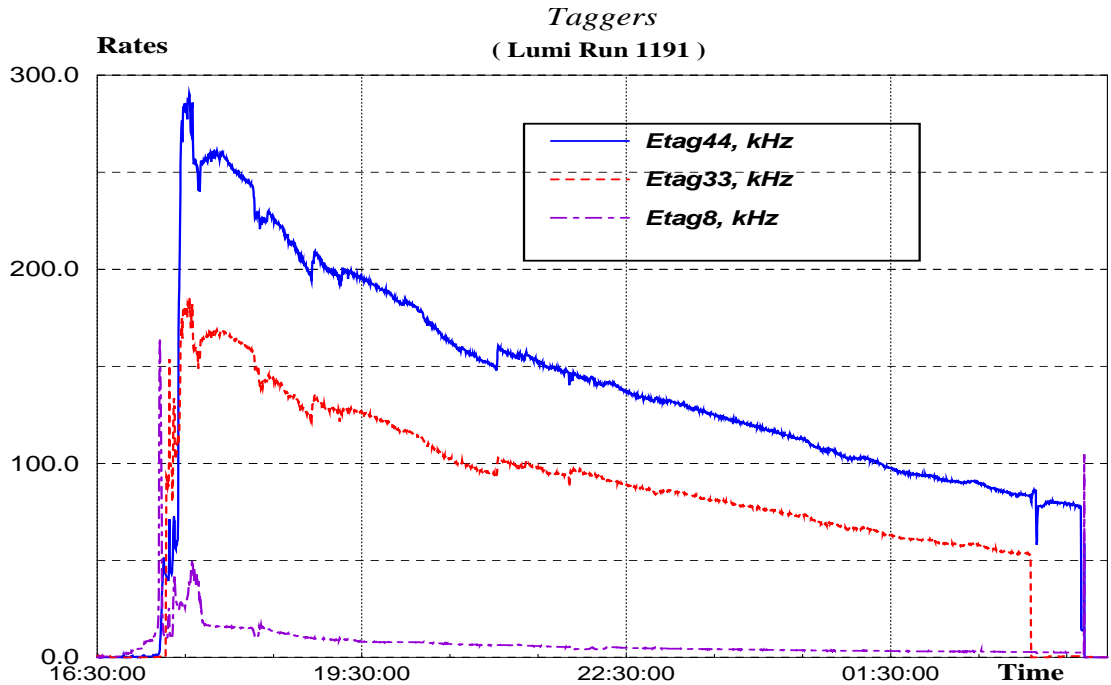,height=150mm,width=90mm,%
    bbllx=20pt,bblly=25pt,bburx=590pt,bbury=770pt,clip=,angle=-90}
\end{picture}
\caption{\footnotesize The behaviour of the 
         rates measured in the electron taggers within
         one beam fill. (Note, ET7 
         labelled ET-8 in figure.)}
\label{fig013}
\end{figure}

The correlations between the rates of the different taggers
within one positron beam filling 
are presented in figure~\ref{fig013}. 
 %-------------------------- fig.13----------- 
%---------------------------------------------------
The measured rates in the 
taggers are in good agreement with 
Monte Carlo simulations of 
bremsstrahlung events including the acceptances 
of the various taggers.  
 
\section{Conclusions}

The addition of a new electromagnetic 
calorimeter, the ET8,
to the H1 detector will 
allow efficient triggering on and 
measurement of high $y$ photoproduction 
events.
A prototype detector, based on a spaghetti type calorimeter
and placed $7\,$m downstream from
the H1 interaction point in the electron direction ($z\approx -7\,$m),
functioned as expected; adequate
energy and position resolution were obtained and 
the trigger performed well.
Following the installation
of a modified section of beam-pipe around  $z\approx -8\,$m, the
ET8 has been installed and will
allow the study of
photoproduction at the highest energies yet achieved
during the HERA 1998 and 1999 running. 

\section{Acknowledgements}

We gratefully acknowledge the support of the DESY 
directorate and the directorate 
and machine group of the LPI synchrotron 
(Moscow). We also wish to thank the 
DESY directorate for the kind hospitality extended to us. 
We are grateful 
to P.\ Biddulph and R.\ Eichler for their
active support and helpful discussions.
We would like to thank INTAS, RFBR and the UK PPARC
for financial support.

\end{document}